\documentclass[a4paper,fleqn,usenatbib]{mnras}

\usepackage[T1]{fontenc}
\usepackage{ae,aecompl}

\usepackage{graphicx}	% Including figure files
\usepackage{amsmath}	% Advanced maths commands
\usepackage{amssymb}	% Extra maths symbols
\usepackage{multirow}

\title[Multiplicity of Galactic Cepheids]{Multiplicity of Galactic Cepheids from long-baseline interferometry~III.\\Sub-percent limits on the relative brightness of a close companion of $\delta$~Cephei}

\author[A.~Gallenne et al.]{A.~Gallenne$^{1}$, A.~M\'erand$^{1}$, P.~Kervella$^{2, 3}$, J.~D.~Monnier$^{4}$, G.~H.~Schaefer$^{5}$,
\newauthor R.~M.~Roettenbacher$^{4}$, W.~Gieren$^{6}$, G.~Pietrzy\'nski$^{6,7}$, H.~McAlister$^{5}$, T.~ten~Brummelaar$^{5}$,
\newauthor J.~Sturmann$^{5}$, L.~Sturmann$^{5}$, N.~Turner$^{5}$
and R.~I.~Anderson$^{8}$
\\
$^{1}$European Southern Observatory, Alonso de C\'ordova 3107, Casilla 19001, Santiago 19, Chile\\
$^{2}$LESIA (UMR 8109), Observatoire de Paris, PSL, CNRS, UPMC, Univ. Paris-Diderot, 5 place Jules Janssen, 92195 Meudon, France\\
$^{3}$Unidad Mixta Internacional Franco-Chilena de Astronom\'{i}a (CNRS UMI 3386), Departamento de Astronom\'{i}a, Universidad de Chile, Camino El Observatorio 1515, Las Condes, Santiago, Chile\\
$^{4}$Department of Astronomy, University of Michigan, 918 Dennison Building, Ann Arbor, MI 48109-1090, USA\\
$^{5}$The CHARA Array of Georgia State University, Mount Wilson CA 91023, USA\\
$^{6}$Universidad de Concepci\'on, Departamento de Astronom\'ia, Casilla 160-C, Concepci\'on, Chile\\
$^{7}$Nicolaus Copernicus Astronomical Centre, Polish Academy of Sciences,  Bartycka 18, 00-716 Warszawa, Poland\\
$^{8}$Department of Physics and Astronomy, Johns Hopkins University, Baltimore, MD 21218, USA
}

\date{Accepted 2016 June 2; Received 2016 June 2; in original form 2016 May 19}

\pubyear{2016}

\begin{document}
\label{firstpage}
\pagerange{\pageref{firstpage}--\pageref{lastpage}}
\maketitle

\begin{abstract}
We report new CHARA/MIRC interferometric observations of the Cepheid archetype $\delta$~Cep, which aimed at detecting the newly discovered spectroscopic companion. We reached a maximum dynamic range $\Delta H $ = 6.4, 5.8, and 5.2\,mag, respectively within the relative distance to the Cepheid $r < 25$\,mas, $25 < r < 50$\,mas and $50 < r < 100$\,mas. Our observations did not show strong evidence of a companion. We have a marginal detection at $3\sigma$ with a flux ratio of 0.21\,\%, but nothing convincing as we found other possible probable locations. We ruled out the presence of companion with a spectral type earlier than F0V, A1V and B9V, respectively for the previously cited ranges $r$. From our estimated sensitivity limits and the Cepheid light curve, we derived lower-limit magnitudes in the $H$ band for this possible companion to be $H_\mathrm{comp} > 9.15, 8.31$ and 7.77\,mag, respectively for $r < 25$\,mas, $25 < r < 50$\,mas and $50 < r < 100$\,mas. We also found that to be consistent with the predicted orbital period, the companion has to be located at a projected separation $< 24$\,mas with a spectral type later than a F0V star.
\end{abstract}

\begin{keywords}
techniques: interferometric -- techniques: high angular resolution -- stars: variables: Cepheids -- star: binaries: close
\end{keywords}

%%%%%%%%%%%%%%%%%%%%%%%%%%%%%%%%%%%%%%%%%%%%%%%%%%

%%%%%%%%%%%%%%%%% BODY OF PAPER %%%%%%%%%%%%%%%%%%

\section{Introduction}

\begin{figure*}
	\centering
	\resizebox{\hsize}{!}{\includegraphics[width = \linewidth]{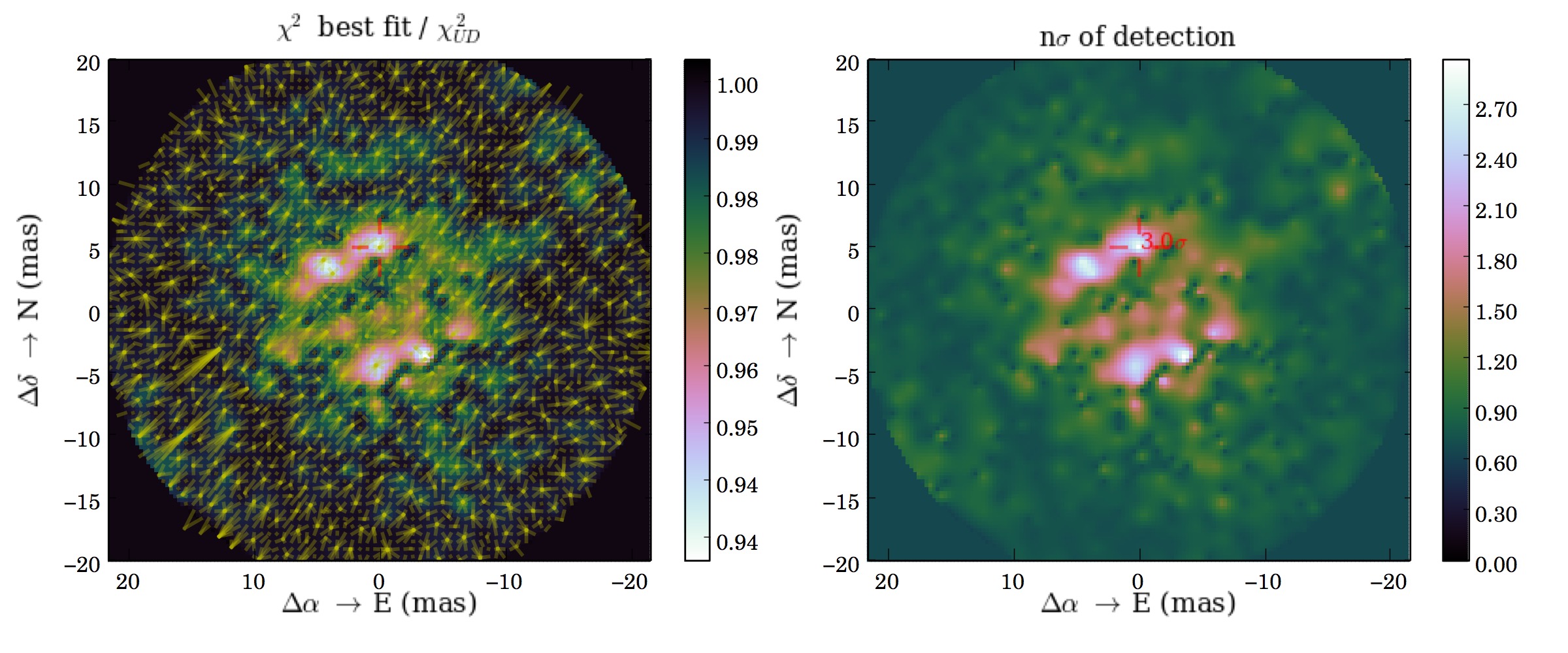}}
	\caption{$\chi_r^2$ map of the local minima (left) and detection level map (right) of $\delta$~Cep using only the closure phase for the observations made in 2015 July. The yellow lines represent the convergence from the starting points to the final fitted position \citep[for more details see][]{Gallenne_2015_07_0}. The maps were re-interpolated in a regular grid for clarity.}
	\label{image__chi2map}
\end{figure*}

$\delta$~Cep is well known as the prototype of classical Cepheid stars since the discovery of its variability by J. Goodricke in 1784 \citep[][although the very first discovered Cepheid was $\eta$~Aql, identified by E. Piggot a few months earlier]{Goodricke_1786__0}. $\delta$~Cep has therefore a particular historical interest as being the Cepheid archetype, and is also an important calibrator for the period-luminosity relation with the most accurate parallax for a Milky Way Cepheid \citep[$272 \pm 11$\,pc,][]{Benedict_2002_09_0}. $\delta$~Cep is also a member of a star cluster \citep{Majaess_2012_03_0}, and its very precise distance derived from cluster membership confirms the parallax distance with a comparable uncertainty (4\,\%). Both distance determinations further show excellent agreement with the value derived by \citet{Storm_2011_10_0} using the infrared surface-brightness technique.

This 5.37\,days pulsating star, which is also the second nearest Cepheid, has been extensively studied with several observing techniques and wavelengths, revealing little by little new unseen physical properties. $\delta$~Cep is known to have an A0-type visual companion located at about 40\arcsec \citep{Fernie_1966_03_0,Prugniel_2007_04_0}, which turns out to have itself an astrometric component \citep{Benedict_2002_09_0}. The  $\delta$~Cep system is also associated with a circumstellar infrared nebulae, reminiscent of a bow shock aligned with the direction of the proper motion of the stars. This might have been created by the interaction between the stellar wind and the local interstellar medium, and might show that the Cepheid is losing mass \citep{Marengo_2010_12_0,Matthews_2012_01_0}. 

Recently, from high-precision radial velocity measurements, \citet{Anderson_2015_05_0} reported the detection of a spectroscopic companion, closer to the Cepheid than the visual component. They estimated an orbital period of about six years, and a projected semi-major axis of 21.2\,mas ($\sim 5.8$\,AU, assuming masses for the components). \citet{Evans_1992_01_0} set upper limits on any possible companions using IUE spectra, she excluded spectral types earlier than A3V ($\sim 2.5\,M_\odot$). \citet{Anderson_2015_05_0} further pointed out that the companion mass is likely to be $< 1.75\,M_\odot$. If the orbit is favorable and the contrast not too high between the components, the companion should be detected by interferometry. We therefore recently performed new and unique multi-telescope interferometric observations with the MIRC instrument at the CHARA array with the goal of detecting this companion. Our team has already proven that long-baseline interferometry is a powerful tool to spatially detect and resolve very close, faint companions orbiting classical Cepheids \citep{Gallenne_2015_07_0,Gallenne_2014_01_0,Gallenne_2013_02_0,Gallenne_2013_04_0}.

\section{CHARA/MIRC observations and data reduction}
\label{section__observation_and_data_reduction}

\begin{table}
	\centering
	\caption{Journal of the observations. $N_{V^2}$ and $N_{CP}$: number of squared visibilities and closure phases.}
	\begin{tabular}{cccc} 
		\hline
		\hline
		UT 						 & MJD						& $N_{V^2}$,	$N_{CP}$	&	Configuration		\\
		\hline
		2015~Jul.~31	&	57234.484		&	508, 379	&	S2-E1-E2-W1-W2			\\
		2015~Oct.~22	&	57317.313		&  303, 160 	&	S2-E1-E2-W1-W2		\\
		\hline
	\end{tabular}
	\label{table__journal}
\end{table}

The observations were performed in July and October 2015 using the Michigan InfraRed Combiner (MIRC) installed at the CHARA array \citep{ten-Brummelaar_2005_07_0}, located on Mount Wilson, California. The CHARA array consists of six 1\,m aperture telescopes with an Y-shaped configuration (two telescopes on each branch), oriented to the east (E1, E2), west (W1,W2) and south (S1, S2), providing good coverage of the $(u, v)$ plane. The baselines range from 34\,m to 331\,m, providing a high angular resolution down to 0.5\,mas in $H$. The MIRC instrument \citep{Monnier_2004_10_0,Monnier_2010_07_0} is an image-plane combiner which enables us to combine the light coming from all six telescopes in $H$. MIRC also offers three spectral resolutions ($R = 42, 150$ and 400), which provide 15 visibility and 20 closure phase measurements across a range of spectral channels.

We observed $\delta$~Cep (HD~213306, HIP~110991) with five telescopes (S2-E1-E2-W1-W2). We used the lowest spectral resolution, where the light is split into eight spectral channels. Table~\ref{table__journal} lists the journal of our observations. We followed a standard procedure of observing a calibrator before and/or after our Cepheids to monitor the interferometric transfer function. The calibrators, HD~206349 ($\theta_\mathrm{UD} = 0.865 \pm 0.011$\,mas) and HD~214454 ($\theta_\mathrm{UD} = 0.593 \pm 0.042$\,mas), were selected using the \textit{SearchCal}\footnote{Available at http://www.jmmc.fr/searchcal.} software \citep{Bonneau_2006_09_0} provided by the Jean-Marie Mariotti Center\footnote{http://www.jmmc.fr}.

The data were reduced using the standard MIRC pipeline \citep{Monnier_2007_07_0}, which consists of computing the squared visibilities and triple products for each baseline and spectral channel, and to correct for photon and readout noises. Squared visibilities are estimated using Fourier transforms, while the triple products are evaluated from the amplitudes and phases between three baselines forming a closed triangle. We then did an incoherent average of 30\,s for the final data such that the projected baselines do not change significantly during one exposure, therefore reducing smearing of the closure phases and optimizing the companion detection sensitivity.

\section{Companion search and sensitivity limit}
\label{section__companion_search_and_sensitivity_limit}

\begin{table*}[!ht]
	\centering
	\caption{$3\sigma$ contrast limits $\Delta m$ of the companion in the $H$ band. The relative distance $r$ is expressed in mas. \#1 and \#2 denote the July and October observations, respectively.}
	\begin{tabular}{c|c|c|c|c|c|c|c|c|c} 
		\hline
		\hline
		&	\multicolumn{3}{c|}{All Observables}	&	\multicolumn{3}{c|}{Only $CP$} 									&	\multicolumn{3}{c}{Sp. Type limit}	 \\
		%		&	\multicolumn{2}{c|}{}		&	\multicolumn{2}{c|}{$r < 100$\,mas}							&	upper limit		&	\\
		&	$r < 25$      & $25 < r < 50$ & $50 < r < 100$    	&	$r < 25$      & $25 < r < 50$ & $50 < r < 100$   	&	$r < 25$      & $25 < r < 50$ & $50 < r < 100$   						\\
		\hline
		\#1	&	6.17	&	5.63	&  5.26	&	6.44	& 5.82	& 5.24	&	F0V	&	A1V	&	B9V		\\
		\hline
		\#2	&	4.24	&	3.68	&  3.51		&	4.35	&  3.45	&  2.88	   & B6V	&	B3V	&	B1V			\\
		\hline
	\end{tabular}
	\label{table__limits}
\end{table*}

To search for companion we used the tool \texttt{CANDID}\footnote{Available at \url{https://github.com/amerand/CANDID}} \citep{Gallenne_2015_07_0}, made for this purpose. It allows a systematic search for companions performing a $N\times N$ grid of fit, whose minimum needed grid resolution is estimated a posteriori. The tool delivers the binary parameters, namely the flux ratio $f$ and the astrometric separation ($\Delta \alpha, \Delta \delta$), but also the uniform disk angular diameter of the primary $\theta_\mathrm{UD}$, and the (non-)detection level of the component. It uses $\chi^2$ statistics to estimate the level of detection in "number of sigmas", and therefore assumes the error bars are uncorrelated. We will claim a detection if the level is $> 3\sigma$.

\begin{figure}
	\centering
	\resizebox{\hsize}{!}{\includegraphics[width = \linewidth]{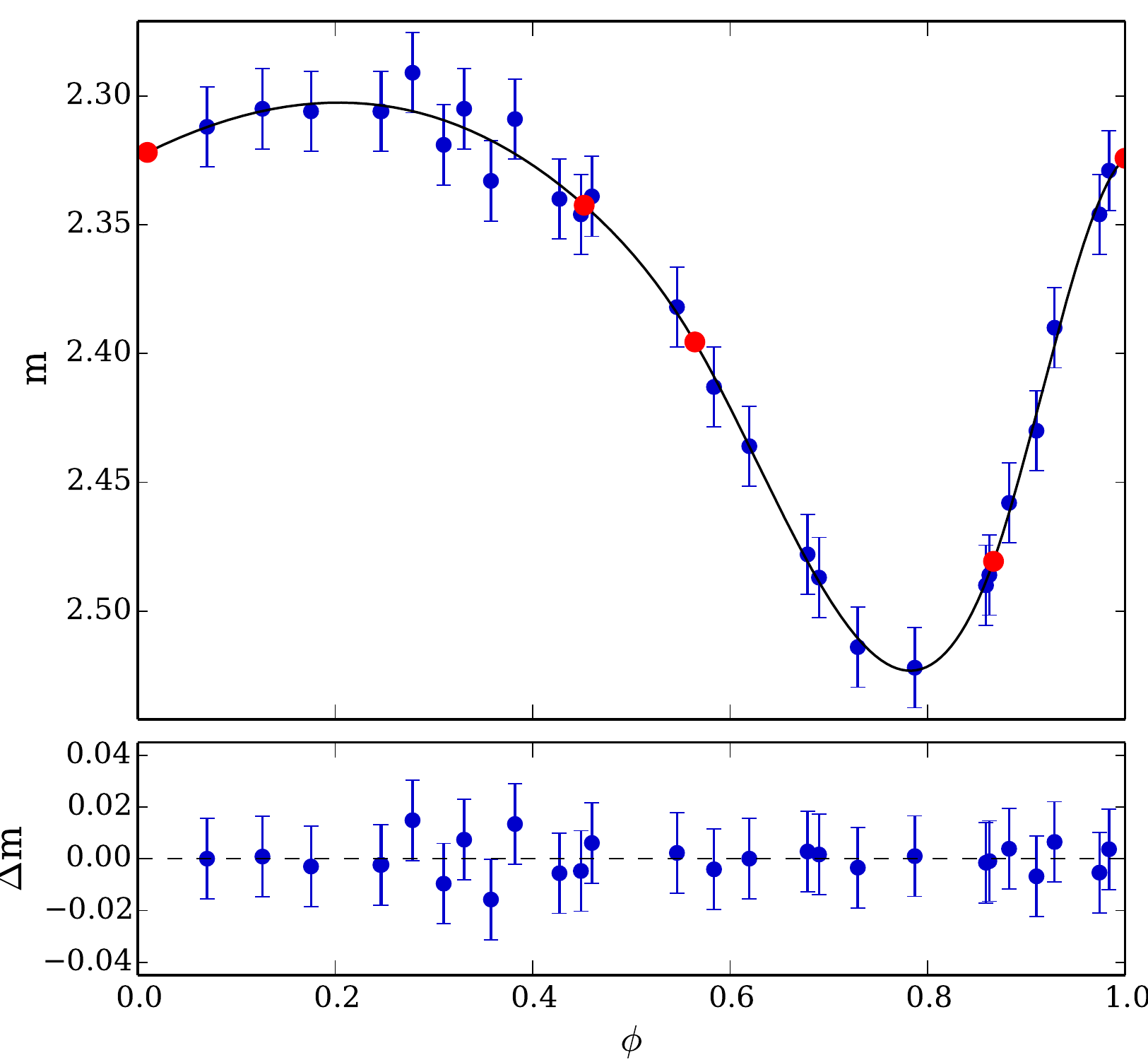}}
	\caption{$H$-band light curve of $\delta$~Cep. The solid line is the periodic cubic-spline function defined with five adjustable floating nodes (red dots).}
	\label{image__lc_curve}
\end{figure}

We first \texttt{CANDID} for each individual dataset (i.e. July and October) and searched around 100\,mas from the Cepheid. We chose this limit because companions at larger distances are strongly impacted by the bandwidth smearing effect, and in addition are more efficiently detected using adaptive optics on a single-dish telescope (through imaging or sparse aperture masking). The first dataset (July) gives a best fit at a detection level of $2.5\sigma$ using all interferometric observables (i.e. the squared visibilities $V^2$, the closure phases CPs, and the bispectrum amplitudes), and $3\sigma$ only using $CPs$. Fig.~\ref{image__chi2map} shows the $\chi_r^2$ map with the most probable location of a companion, if any, and the $n\sigma$ map giving the detection level at each point in the grid for the July observations. A companion might be detected at $\rho \sim 5$\,mas and $PA \sim -0.8^\circ$, with a flux ratio of $f \sim 0.21$\,\%, but other positions seem also possible with similar detection levels. Furthermore, such a flux ratio is below the average sensitivity limit reachable by the current beam combiners, but two components were already detected with such a contrast \citep{Roettenbacher_2015_08_0,Gallenne_2015_07_0}. The second dataset (October) gives a best fit at a detection level of $3.1\sigma$ both for all observables and only the CPs, but at distinct locations and very different flux ratios. As this second observation was performed under poor seeing conditions, the visibilities are probably biased and not reliable to detect such faint companions. With the closure phase only, the most probable location is at $\rho \sim 7$\,mas and $PA \sim 2^\circ$, and seems consistent with the July observations. However, the estimated flux ratio is 0.88\,\%, very different to 0.21\,\% in July. The magnitude difference of the Cepheid between the two phases is $\sim +0.03$\,mag (see Fig.~\ref{image__lc_curve}), which would correspond to a $\sim -3$\,\% change in flux ratio, and not a factor of 4. We therefore conclude that we do not have a detection in October, while it is marginal for July and just at our chosen detection threshold. %To increase the signal-to-noise ratio, we ran \texttt{CANDID} combining the two datasets, but only using the CPs. This is relevant because the closure phase is insensitive to the angular diameter variation of the Cepheid, but prevents any detection of short orbital period. However, the expected period is about six years \citep{Anderson_2015_05_0}, which means that the possible companion would have moved of only 4\,\% between our two datasets. This combination gives a maximum detection level at $1.4\sigma$, also leading to a non-detection of the companion.

\texttt{CANDID} has also implemented a robust method to derive the dynamic range we can reach with a given set of data. It consist of injecting a fake companion into the data at each astrometric position with different flux ratios. As we inject a companion, we therefore know that the binary model should be the true model. We then compare the $\chi^2$ with the one of a single star model (uniform disk model) to obtain the probability of the binary model to be the true model. We set the significance level on the flux ratios at $3\sigma$, meaning that lower flux ratios are not significantly detected. We refer the reader to \citet{Gallenne_2015_07_0} for more information about the method. For comparison, \texttt{CANDID} has also implemented a less robust method which consist of comparing a uniform disk model with a binary model for each position in the grid, and then check whether the probability of the binary model is consistent with the data \citep{Absil_2011_11_0}. In the following, all of the given detection limits are derived from our injection method, because we demonstrated in \citet{Gallenne_2015_07_0} that the Absil's method may under- or overestimate the detection limits.

\begin{figure}
	\centering
	\resizebox{\hsize}{!}{\includegraphics[width = \linewidth]{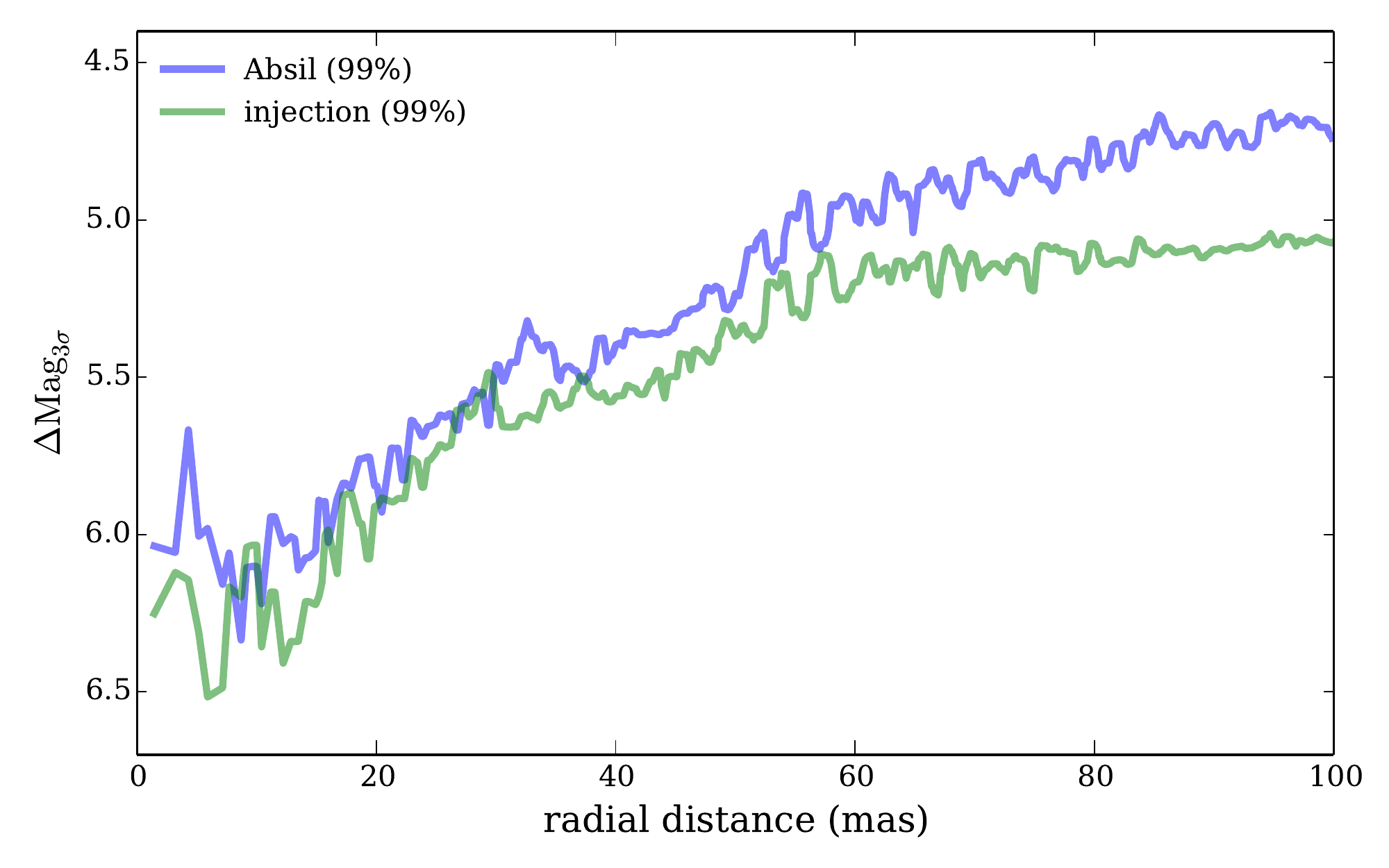}}
	\caption{Contrast limit at $3\sigma$, $\Delta m_{3\sigma}$, for a companion orbiting $\delta$~Cep.}
	\label{image__limit}
\end{figure}

We listed in Table~\ref{table__limits} two detection limits for each dataset, one using all of the observables, and one using only the closure phases. We also gave three different values, the average for $r < 25$\,mas, $25 < r < 50$\,mas and $50 < r < 100$\,mas, which can be relevant when the limit increases with the relative distance to the Cepheid, $r$. All of the final $3\sigma$ contrast limits, $\Delta m_{3\sigma}$ expressed in magnitude, are conservative as they correspond to the mean plus the standard deviation for the given radius range. We present in Fig.~\ref{image__limit} the contrast limit curve, using all observables, for the observations performed in July.

\begin{figure}
	\centering
	\resizebox{\hsize}{!}{\includegraphics[width = \linewidth]{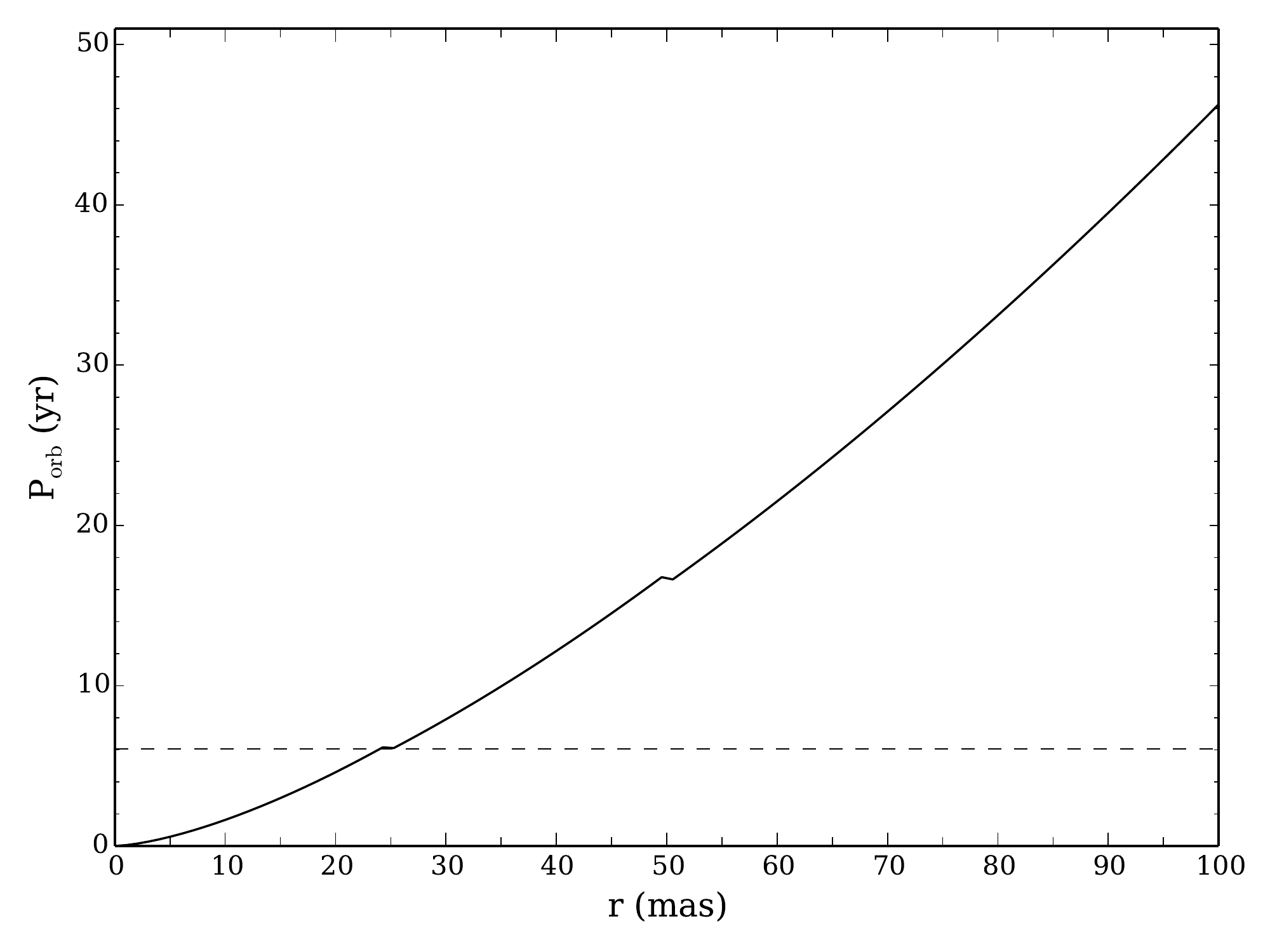}}
	\caption{Minimum possible orbital period in function of the projected radial distance. The dashed line represents the period estimated from \citet{Anderson_2015_05_0}.}
	\label{image__period_limit}
\end{figure}

From an evolutionary timescale point of view, most of the companions should be stars close to the main sequence. We therefore set upper limits for the companion spectral type assuming it is on the main sequence, and based on their H-band luminosities. From our estimated reported limits, we can exclude the presence of a companion having a spectral type earlier than a F0V star within 25\,mas, A1V star between 25--50\,mas, and B9V star between 50--100\,mas from $\delta$~Cep (estimated using intrinsic colors from \citet{Cox_2000__0} and \citet{Ducati_2001_09_0}). This would correspond to a companion mass $< 1.6\,M_\odot$, $< 2.7\,M_\odot$, and $< 3.4\,M_\odot$, respectively \citep{Cox_2000__0}. From those limits, we can check the consistency with the predicted orbit of \citet{Anderson_2015_05_0}. Using Kepler's law and setting the projected separation $r$ as a lower limit for the angular semi-major axis, that is $a \geq r$, we have:
\begin{displaymath}
P^2 \geq \frac{r^3 d^3}{M_1 + M_2},
\end{displaymath}
with $r$ in arcsecond, $d$ in parsec, $P$ in year and the masses in solar mass. In Fig.~\ref{image__period_limit}, we plotted the function $P_\mathrm{min}(r)$, using a maximum Cepheid mass of $6\,M_\odot$ \citep[see e.g.][]{Matthews_2012_01_0}, and the previously cited mass limits for the companion. We also plotted the predicted upper-limit period (i.e. $P + \sigma$) derived by \citet{Anderson_2015_05_0}. We can see that to be consistent with the expected period, the companion has to be located at $r < 24$\,mas, and has therefore a spectral type later than a F0V star. It is also consistent with the predicted projected semi-major axis of 21.2\,mas.

The marginally detected component with $f = 0.21$\,\% ($\Delta m = 6.7$\,mag) would correspond to a F2V star, with a mass of about $1.5\,M_\odot$, which would be consistent with the $1.75\,M_\odot$ upper limit pointed out by \citet{Anderson_2015_05_0}. The projected separation of $\sim 5$\,mas seems also consistent with our previous maximum separation threshold.

We can also set lower limit on the companion flux using the Cepheid $H$-band light curve from the literature, which would actually represent the combined flux in case of an unseen companion is present. We retrieved the photometry from \citet{Barnes_1997_06_0} and used the ephemeris $T_0$ = 2448305.2362421\,days and $P_\mathrm{puls}$ = 5.3662906\,days \citep{Merand_2015_12_0} to construct the $H$-band light curve. To estimate the magnitudes at our given pulsation phases (i.e. $\phi$ = 0.05 and 0.48), we interpolated the data with a periodic cubic-spline function defined by floating nodes. The interpolated curve is shown in Fig.~\ref{image__lc_curve}. We then estimated the combined magnitudes $2.32 \pm 0.01$\,mag and $2.35 \pm 0.01$\,mag, respectively at phases $\phi = 0.05$ and 0.48. For simplicity and as those values are rather close, we used the averaged value and the standard deviation, giving the average combined magnitude $m_{12} = 2.34 \pm 0.02$\,mag. Following Eq.~2 of \citet{Gallenne_2014_01_0}, 
\begin{equation}
m_2 = m_{12} + 2.5\log(1 + 1/f),
\end{equation}
with $\Delta m = -2.5\,\log{f}$, and $f$ the flux ratio between the companion and the Cepheid, we estimated a minimum $H$-band magnitude for the companion to be $H_\mathrm{comp} > 9.15, 8.31$ and 7.77\,mag, respectively for $r < 25$\,mas, $25 < r < 50$\,mas and $50 < r < 100$\,mas.

\begin{figure}
	\centering
	\resizebox{\hsize}{!}{\includegraphics[width = \linewidth]{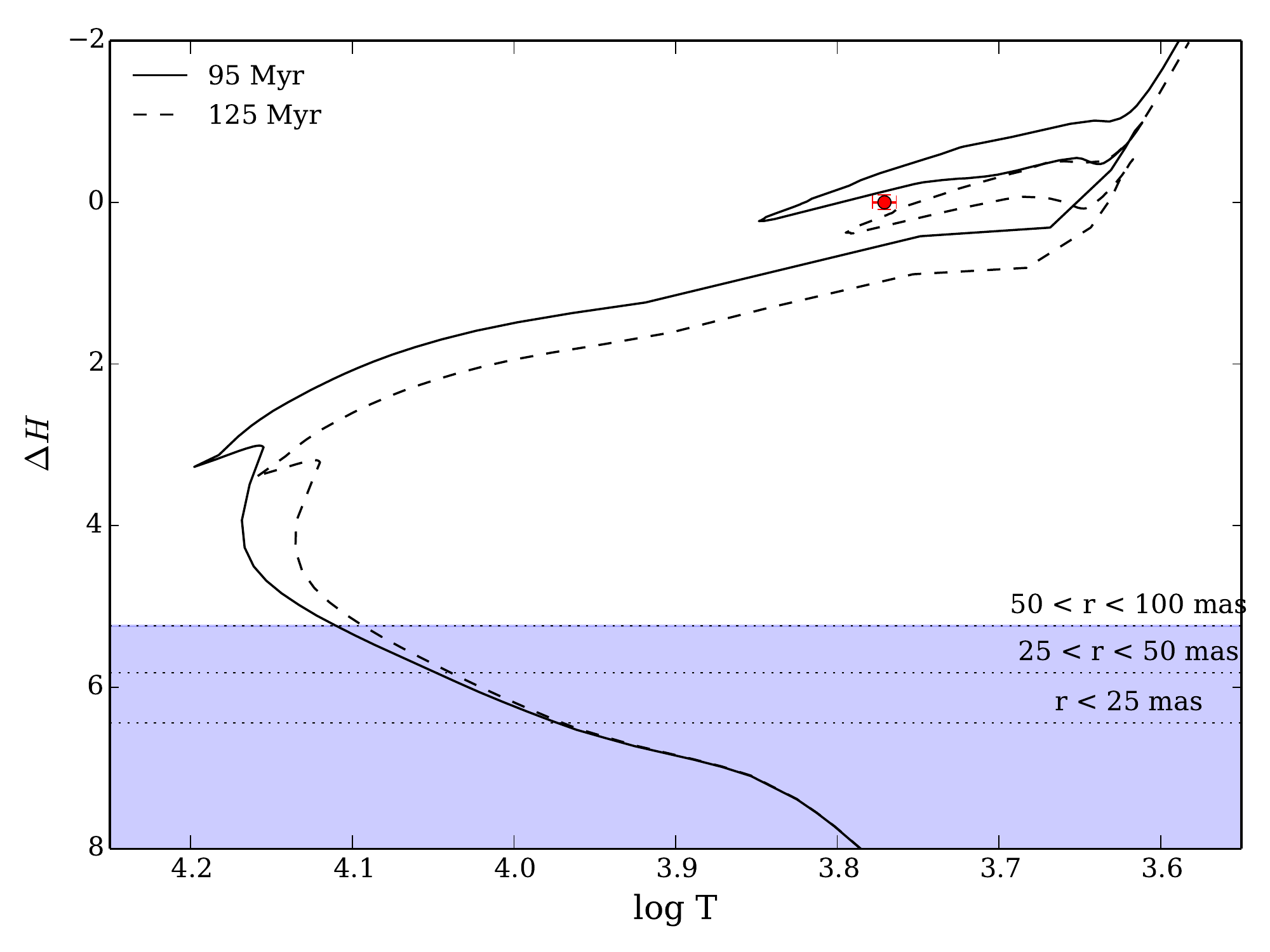}}
	\caption{Isochrones from the Geneva evolution models \citep{Ekstrom_2012_01_0}, normalized to the $H$-band absolute magnitude of the Cepheid. The red dot denotes the position of $\delta$~Cep, while the dashed lines represent our derived magnitude limits. The blue area shows the possible luminosity class for an undetected companion. %Spectral types at corresponding average temperatures are indicated in the upper $x$ axis.
	}
	\label{image__hr}
\end{figure}

Another approach in deriving the luminosity class limit of this possible companion is to use the Hertzsprung-Russell (HR) diagram and our derived magnitude difference limits. We retrieved Geneva stellar isochrones  \citep{Ekstrom_2012_01_0} for ages $t = 95-125$\,Myr, a solar metallicity $Z = 0.014$ and a initial rotation rate of $\Omega / \Omega_\mathrm{crit} = 0.568$. Using the output of the isochrone models, i.e. effective temperatures and absolute magnitudes, we converted the $L-T_\mathrm{eff}$ HR diagram into a $H$-band absolute magnitude-temperature diagram ($\Delta H-T_\mathrm{eff}$) using the known properties of the Cepheid, i.e. the parallax $\pi = 3.66 \pm 0.15$\,mas \citep{Benedict_2002_09_0}, an average effective temperature $T_\mathrm{eff} = 5900 \pm 100$\,K \citep{Andrievsky_2005_10_0}, and an average apparent magnitude $m_\mathrm{H} = 2.38 \pm 0.01$\,mag \citep{Barnes_1997_06_0}. In Fig.~\ref{image__hr}, we show two isochrones encompassing the Cepheid properties (red dot). We see that the two isochrone limits give the same luminosity class limit for an undetected component, as indicated by the blue area in the figure, i.e. if a companion is orbiting $\delta$~Cep, it should be a main-sequence or white dwarf star. The presence of a white-dwarf companion is not unlikely. This is possible if the companion was originally more massive than the Cepheid, but no white dwarf companion to a Cepheid has been found so far.

%We also indicated the spectral types for main-sequence stars corresponding to their average temperature taken from \citet{Cox_2000__0}. Although still consistent with our previous estimated spectral type limits (\ref{table__limits}), the corresponding limits according the isochrones are slightly different. This is due to the calibration of the MK spectral type with temperature which can differ by several hundreds of kelvin.

\section{Angular diameter}

\begin{figure}
	\centering
	\resizebox{\hsize}{!}{\includegraphics[width = \linewidth]{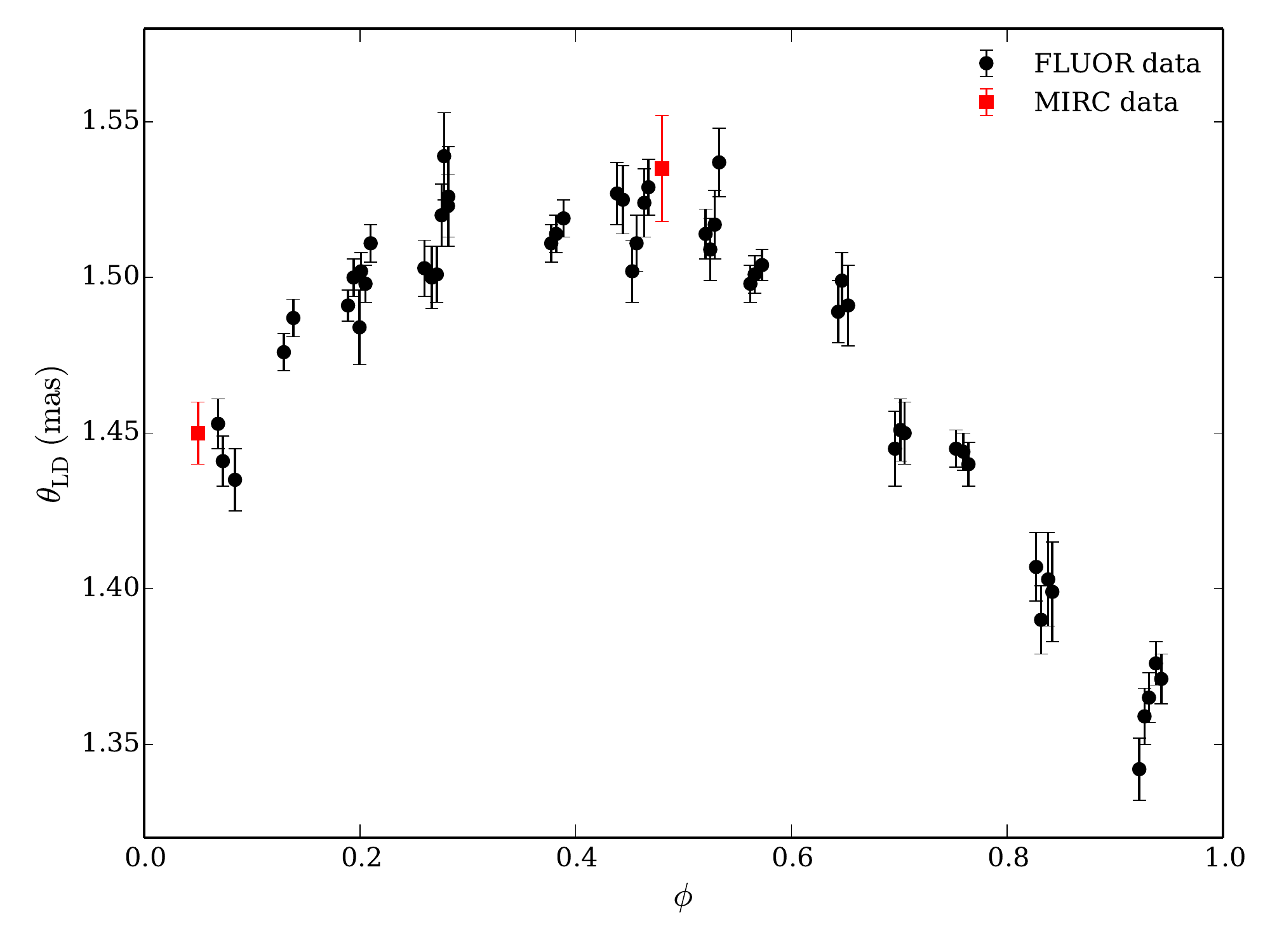}}
	\caption{Limb-darkened disk angular-diameter variation of $\delta$~Cep.
	}
	\label{image__ld}
\end{figure}

Our observations also provide measurements of the angular diameter of the Cepheid for two pulsation phases. We estimated the limb-darkened (Rosseland) diameter of the star following the formalism of \citep{Merand_2015_12_0}, i.e. we extracted the radial intensity profile $I(r)$ of the spherical SATLAS models (for temperatures of 6900\,K and 5600\,K at phases 0.05 and 0.048), which was converted to a visibility profile using a Hankel transform and fitted to our squared visibility data.

We measured $\theta_\mathrm{LD} = 1.450 \pm 0.010$\,mas and $\theta_\mathrm{LD} = 1.535 \pm 0.017$\,mas, respectively at pulsation phases 0.05 and 0.48. These values have been estimated using the bootstrapping technique (with replacement) on all baselines. We took from the distributions the median, and the maximum value between the 16th and 84th percentiles as uncertainty. These measurements are in good agreement with the angular diameter variation curve previously reported with the instrument CHARA/FLUOR \citep{Merand_2005_07_0,Merand_2015_12_0}, as shown in Fig.~\ref{image__ld}. 

\section{Conclusion}
\label{section__conclusion}

We reported new and unique multi-telescope interferometric observations of the classical Cepheid $\delta$~Cep. The goal was to detect the newly discovered spectroscopic companion reported by \citet{Anderson_2015_05_0}.

Our observations did not show strong evidence of any companion with a spectral type earlier than F0V, A1V and B9V, respectively within the relative distance from $\delta$~Cep $r < 25$\,mas, $25 < r < 50$\,mas and $50 < r < 100$\,mas. The spectral type limits are tighter than previous works for $r < 25$\,mas. We also estimated lower limit magnitudes for the companion to be $H_\mathrm{comp} > 9.15, 8.31$ and 7.77\,mag, respectively. We also showed that to be consistent with the predicted orbital period, the companion has to be located at a projected separation $< 24$\,mas and have spectral type later than a F0V star.

A component might have been marginally detected at only $3\sigma$, but we found several possible locations. Although the estimated flux ratio and separation are consistent with what expected, the detection is not really convincing. Additional data are necessary to claim a detection.

The regular dynamic range reachable by the current beam combiners is about 5.8\,mag (200:1), making this new possible companion of $\delta$~Cep hardly detectable from long-baseline interferometry, but not impossible as already demonstrated by \citet{Gallenne_2015_07_0} and \citet{Roettenbacher_2015_08_0}, who detected components with a flux ratio of about 6.5\,mag.

\section*{Acknowledgements}
The authors would like to thank the CHARA Array and Mount Wilson Observatory staffs for their support. This work is based upon observations obtained with the GSU Center for High Angular Resolution Astronomy Array at Mount Wilson Observatory. The CHARA Array is supported by the NSF under Grants No. AST-1211929 and AST-1411654. Institutional support has been provided from the GSU College of Arts and Sciences and the GSU Office of the Vice President for Research and Economic Development. JDM's contribution to this work was partially supported by HST Guest Observer grants (HST-GO-13454.07-A, HST-GO-13841.006-A) and  NSF (NSF-AST1108963). WG and GP gratefully acknowledge financial support for this work from the BASAL Centro de Astrof\'isica y Tecnolog\'ias Afines (CATA) PFB-06/2007. Support from the Polish National Science Centre grants MAESTRO DEC-2012/06/A/ST9/00269 is also acknowledged. PK, AG and WG acknowledge support of the French-Chilean exchange program ECOS-Sud/CONICYT (C13U01). The authors acknowledge the support of the French Agence Nationale de la Recherche (ANR), under grant ANR-15-CE31-0012-01 (project UnlockCepheids). R.I.A. acknowledges funding from the Swiss National Science Foundation. W.G. also acknowledges financial support from the Millenium Institute of Astrophysics (MAS) of the Iniciativa Cientifica Milenio del Ministerio de Econom\'ia, Fomento y Turismo de Chile, project IC120009. This research received the support of PHASE, the high angular resolution partnership between ONERA, Observatoire de Paris, CNRS, and University Denis Diderot Paris 7. This work made use of the SIMBAD and VIZIER astrophysical database from CDS, Strasbourg, France and the bibliographic informations from the NASA Astrophysics Data System. This research has made use of the Jean-Marie Mariotti Center \texttt{SearchCal} and \texttt{ASPRO} services, co-developed by FIZEAU and LAOG/IPAG, and of CDS Astronomical Databases SIMBAD and VIZIER.

%%%%%%%%%%%%%%%%%%%%%%%%%%%%%%%%%%%%%%%%%%%%%%%%%%


\bibliographystyle{mnras}   % if natbib is available
\bibliography{./bibliographie}

% Don't change these lines
\bsp	% typesetting comment
\label{lastpage}
\end{document}